\documentclass[a4paper,11pt]{article}
\usepackage{pos}
\usepackage{multirow}
\usepackage{float}
\usepackage{caption}
\usepackage{subcaption}

\newcommand{\amu}{\mbox{$a_\mu^{\mathrm{HVP, LO}}$}}

\title{Calculating the QED correction to the hadronic vacuum polarisation on the lattice}

\author*[a]{Gaurav~Ray}
\author[b]{Alexei~Bazavov}
\author[c]{Christine~Davies}
\author[d]{Carleton~DeTar}
\author[e]{Aida~El-Khadra}
\author[f]{Steven~Gottlieb}
\author[c]{Daniel~Hatton}
\author[f]{Hwancheol~Jeong}
\author[g]{Andreas~Kronfeld}
\author[e]{Shaun Lahert}
\author[i]{Peter~Lepage}
\author[a]{Craig~McNeile}
\author[h]{James~Simone}
\author[d]{Alejandro~Vaquero~Avilés-Casco}

\affiliation[a]{Centre for Mathematical Sciences, University of Plymouth, Plymouth, PL4 8AA, UK}
\affiliation[b]{Department of Physics and Astronomy, Michigan State University, East Lansing, MI 48824, USA}
\affiliation[c]{SUPA, School of Physics and Astronomy, University of Glasgow, Glasgow, G12 8QQ, UK}
\affiliation[d]{Department of Physics and Astronomy, University of Utah, Salt Lake City, UT 84112, USA}
\affiliation[e]{Department of Physics, University of Illinois, Urbana, IL 61801, USA}
\affiliation[f]{Department of Physics, Indiana University, Bloomington, Indiana, 47405, USA}
\affiliation[g]{Fermi National Accelerator Laboratory, Batavia, IL 60510 USA}
\affiliation[i]{Laboratory for Elementary-Particle Physics, Cornell University, Ithaca, NY 14853, USA}

\emailAdd{gaurav.sinharay@plymouth.ac.uk}

\abstract{Isospin-breaking corrections to the hadron vacuum
polarization  component of the
  anomalous magnetic moment of the muon are needed to ensure the
  theoretical precision of $g_\mu -2$ is below the experimental
  precision. We describe the status of our work calculating, using
  lattice QCD, the QED correction to the light and strange
  connected hadronic vacuum polarization in a Dashen scheme. 
We report results using
  physical $N_f=2+1+1$ HISQ ensembles at three lattice 
spacings and three
  heavier-than-light valence quark masses.  }

\FullConference{%
  The 39th International Symposium on Lattice Field Theory (Lattice2022),\\
  8-13 August, 2022 \\
  Bonn, Germany 
}

\begin{document}
\maketitle

\section{Introduction}

The experimental results from the Fermilab Muon g-2  
experiment~\cite{Muong-2:2021ojo} 
and E821 experiment~\cite{Muong-2:2006rrc},
for the muon anomalous magnetic moment,
motivates reducing the errors on lattice QCD calculations of the leading
order hadronic vacuum polarization
contribution to the muon anomalous magnetic moment (\amu).
There is a comprehensive review~\cite{Aoyama:2020ynm}
of the theoretical calculations
of \amu.

This project is part of the Fermilab Lattice, HPQCD and MILC 
collaboration's~\cite{Chakraborty:2014mwa,Chakraborty:2016mwy,FermilabLattice:2017wgj,FermilabLattice:2019ugu,Lahert:2021xxu,FermilabLattice:2022izv} work on computing \amu. 
Reducing the theoretical uncertainty of $a^{\text{HVP,LO}}_\mu$ below 1\% requires the inclusion of isospin breaking effects. These arise from the up and down quarks unequal masses, $m_u\neq m_d$, and their unequal electric charges, $Q_u\!=\!2e/3\neq Q_d\!=\!-e/3$. This project aims to calculate the QED isospin breaking correction to the light and strange connected \amu. 

We use the following definition for the QED correction to the \amu, $\delta a^{f}_{\mu}$,
\begin{equation}
\begin{aligned}
    \delta a^{(f)}_{\mu} &\equiv a^{f}_{\mu}(m_f,Q_f) - a^{f}_{\mu}(m_f,0),
\end{aligned}
\end{equation}
where $f$ labels the quark flavour and the difference is evaluated at equal renormalised quark mass. The QED correction to the connected strange \amu is then
\begin{equation}
    \delta a^{(s)}_{\mu} = a^{s}_{\mu}(m_s,-1/3e) - a^{s}_{\mu}(m_s,0),
    \label{eqn:delta_amu_strange}
\end{equation}
with corresponding formulas for the up and down quarks. The QED correction to the light connected \amu is the sum of the corrections for the up and down quarks,
\begin{equation}
     \delta a^{(l)}_\mu = \delta a^{(u)}_{\mu} + \delta a^{(d)}_{\mu}.
\end{equation}

We originally extract QED corrections to $a_\mu$ at fixed bare quark
mass ($\Delta a_\mu$) and then convert to $\delta a_\mu$ using 
\begin{equation}
    \delta a_\mu = \Delta a_\mu - \delta m_q  \frac{\partial a_\mu}{\partial m_q}
\label{eq:scheme}
\end{equation}
A Dashen-like scheme is used to set the renormalized quark mass following~\cite{MILC:2018ddw,Fodor:2016bgu}.

\section{Simulation Details}

We measured correlators on gauge field ensembles generated with the
Highly Improved Staggered Quark (HISQ) action~\cite{Follana:2006rc}
with 2+1+1 flavours of dynamical sea quarks and physical pion
masses. In order to take the continuum limit we took measurements on
three ensembles with lattice spacings of approximately 0.15, 0.12, and
0.09 fm. The HISQ ensembles were generated by the MILC
collaboration~\cite{MILC:2010pul,MILC:2012znn}. The basic parameters
of the ensembles are in Table~\ref{tab:ensembles}. The lattice spacing
is fixed using the Wilson flow parameter $w_0\!=\!0.1715(9)$
fm~\cite{Dowdall:2013rya}.

\begin{table}[H]
    \centering
    \begin{tabular}{ c c c c c c c } \hline
      name & $L^3$x$T$ & $w_0/a$ & $M_{\pi}L$ & $M_{\pi}$ (MeV) & $N_{\mathrm{cfg}}$  \\ \hline
      very coarse & 32$^3$x48 & 1.13215(35) & 3.30 & 134.73(71) & 1844 \\
      coarse      & 48$^3$x64 & 1.41490(60) & 3.88 & 132.73(70)  & 967 \\
      fine        & 64$^3$x96 & 1.95180(70) & 3.66 & 128.34(68)  & 596 \\
      \hline
    \end{tabular}
    \caption{Properties of the gauge field ensembles used in the
      measurements. The lattice spacings, $w_0/a$,
      and pion masses are from~\cite{FermilabLattice:2019ugu}.}
    \label{tab:ensembles}
\end{table}

 We use the electro-quenched
 approximation~\cite{Duncan:1996xy,MILC:2018ddw,Hatton:2020qhk} to
 partially include the dynamics of QED. The quenched QED fields were
 fixed to the Feynman gauge with the QED$_{L}$
 prescription~\cite{Hayakawa:2008an}.  We measure pseudoscalar and
 vector correlators with equal mass, oppositely charged quarks and
 antiquarks, so that all the mesons are neutral. The code first reads in
 a dynamical SU(3) gauge configuration and a quenched U(1) gauge
 configuration before multiplying the U(1) link fields into the SU(3)
 link fields, and gauge smearing as usual.

We use stochastic-wall sources projected onto the appropriate
spin-taste quantum numbers. For the vector current we use the
$\gamma_i\otimes\gamma_i$ operators, and for the pseudoscalar current
we use the $\gamma_5\otimes\gamma_5$ operator.  The local vector
current is not conserved, therefor it requires renormalising with a
renormalisation factor $Z_V$. The required $Z_V$, including the QED
corrections, have been calculated by
HPQCD~\cite{Hatton:2019gha,Hatton:2020qhk}.  To remove potential
subjective bias, we do a ``blinded analysis''.  The blinding is done by
multiplying the correlators on each ensemble by a random hidden number in [0.95,1.05].

Following BMW~\cite{Borsanyi:2020mff}, in order to avoid the increased
statistical noise incurred by simulating at the light quark mass,
$m_l\equiv 1/2(m_u+m_d)$, we measure with valence quarks at multiples
of $m_l$. We measure at $3 m_l$, $5 m_l$, and $7 m_l$
and the strange quark mass, $m_s$, on each ensemble. We
use a multi-shift solver so that for each charge all the masses can be
solved in a single iterative process. As it is not prohibitively
expensive we also measure at the physical $m_u$ and $m_d$ on the very
coarse ensemble. The correlators measured, including the quark masses
and electric charges, are summarised in
Table~\ref{tab:valence_masses}.

\begin{table}[H]
    \centering
    \begin{tabular}{c c c c } \hline
      name & Charges ($e$) & Quark masses ($am_q$) & sources\\ \hline
       \multirow{2}*{very coarse} & \multirow{2}*{ $\pm 2/3$,$\pm 1/3$,0 } &  0.001524, 0.003328, &  \multirow{2}*{16} \\ & & 0.007278, 0.01213, 0.01698, 0.0677 & \\ 
      coarse       & $\pm 2/3$, $\pm 1/3$, 0 & 0.00552, 0.0092, 0.01288, 0.0527 & 16  \\
      fine        & $\pm 2/3$, $\pm 1/3$, 0 & 0.0036, 0.006, 0.0084, 0.0364 & 16  \\
      \hline
    \end{tabular}
    \caption{The valence quark masses and charges used to compute the pseudoscalar and vector correlators.}
    \label{tab:valence_masses}
\end{table}
We measure three sets of correlators, one uncharged where the SU(3) link fields are not multiplied by U(1) fields, and two charged with opposite electric charges. All the correlators are overall electrically neutral. We calculate two sets of correlators in this way as the charged correlators are noisier than the uncharged correlators. This is due to the presence of a QED noise term proportional to the electric charge, $e$, in the propagator. To suppress this noise term we average over the two correlators with opposite charges. 
 
We use 16 time-sources on each field configuration to improve the
statistics. All other things being equal this would increase the
resources needed to run the simulations by a factor of 16. To mitigate
this we use the truncated solver method
(TSM)~\cite{Bali:2009hu}\cite{MILC:2010pul}. We use 16 sloppy solves with a
residual of $10^{-3}$ and 1 precise solve with a residual of $10^{-6}$
before averaging over all the solves using the TSM method.

\section{Results}

\begin{figure}
     \centering
     \begin{subfigure}[t]{0.4\textwidth}
         \centering
         \includegraphics[scale=0.4]{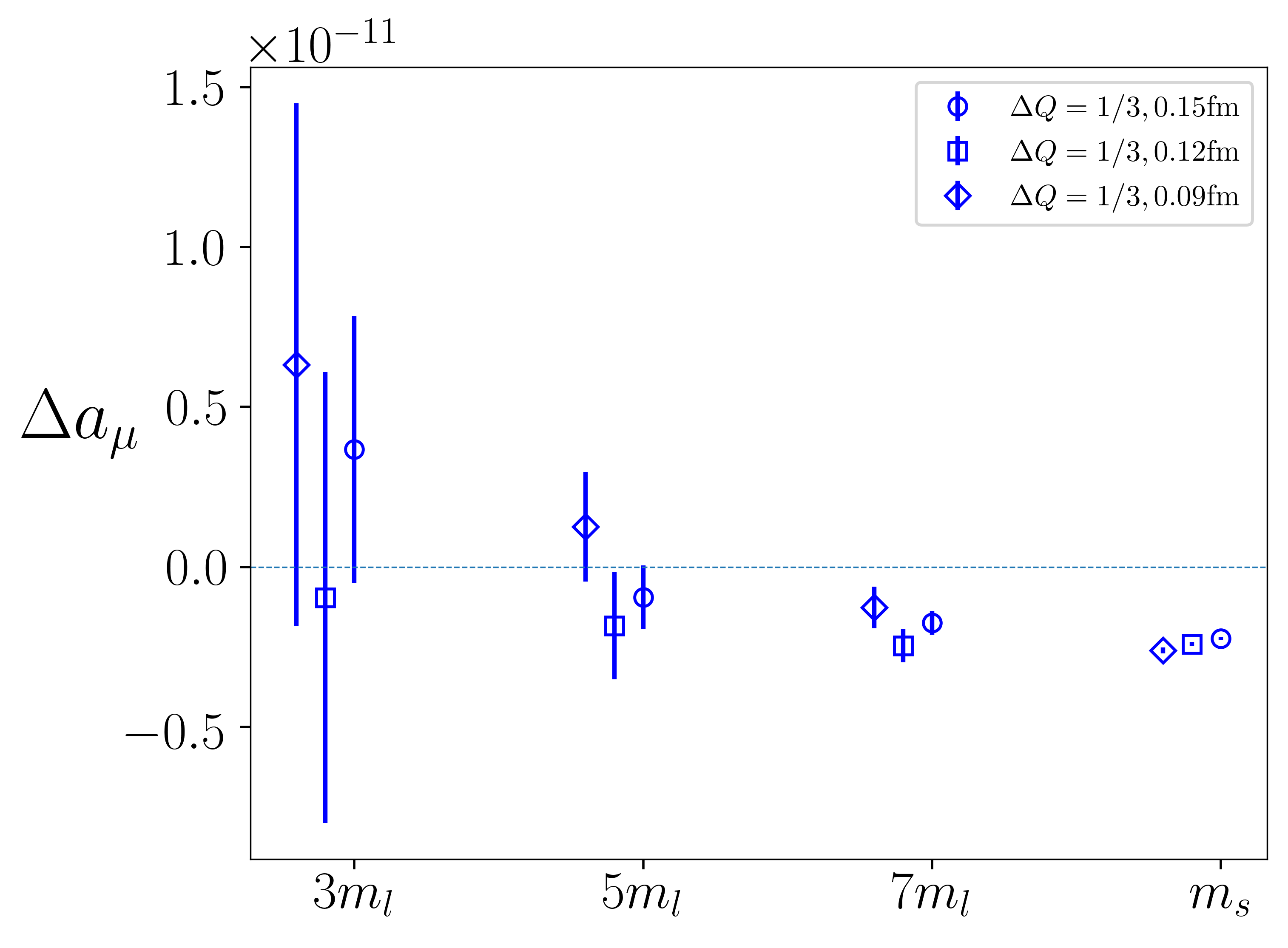}
         \caption{$\Delta a_\mu$ for down-like ($Q\!=\!-1/3e$)
 electric charges.
         \label{fig:corrDiff}}
     \end{subfigure}
     \hfill
     \begin{subfigure}[t]{0.4\textwidth}
         \centering
         \includegraphics[scale=0.4]{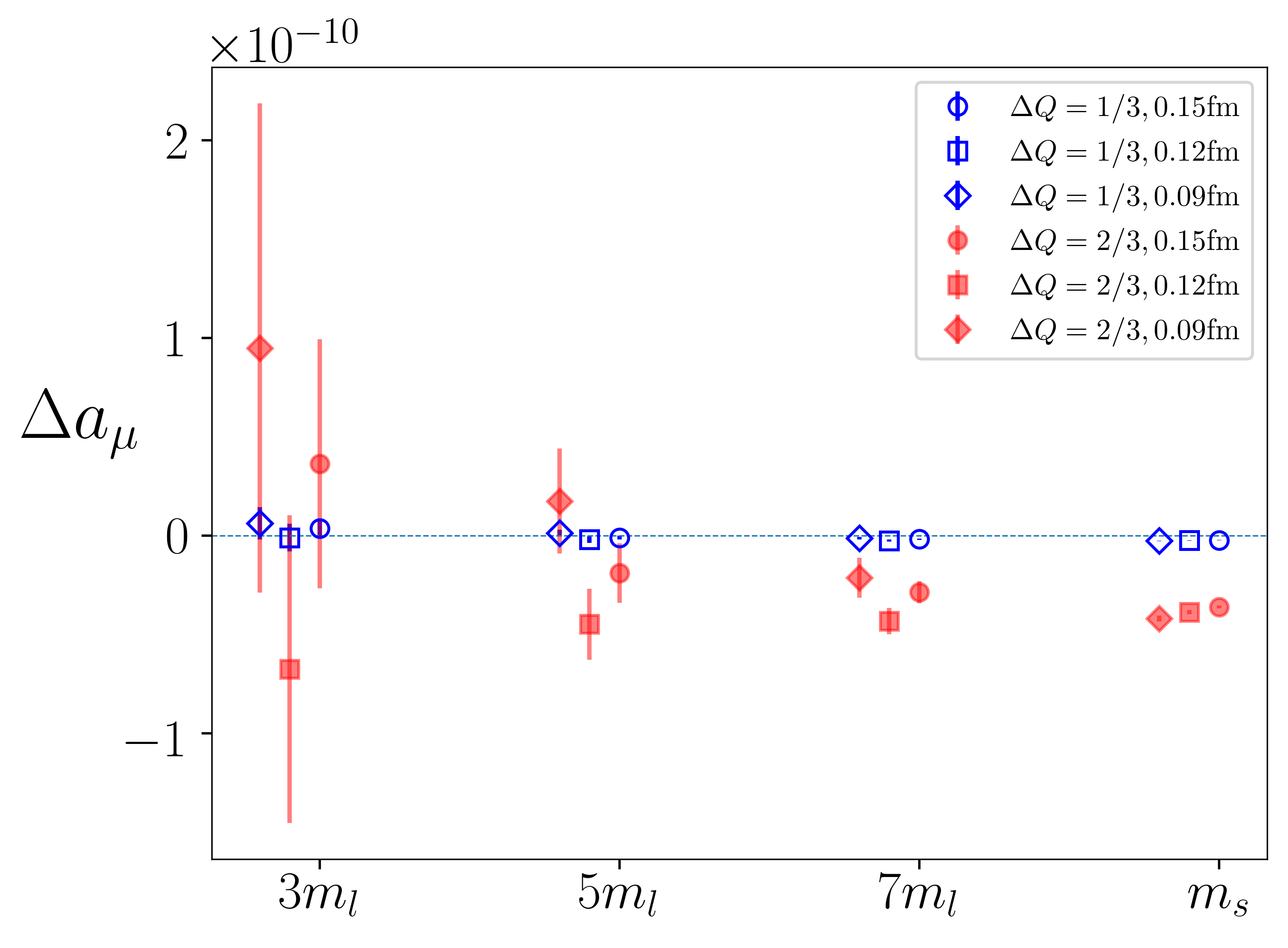}
         \caption{$\Delta a_\mu$ for down-like ($Q\!=\!-1/3e$) 
and up-like ($Q\!=\!2/3e$) electric charges.
         \label{fig:bottomDiff}}
     \end{subfigure}
\caption{The blinded QED corrections to \amu for all the
  ensembles and quark masses at fixed bare quark mass.
\label{fig:chargeDiff}}
 \end{figure}

The QED corrections at equal bare quark mass, $\Delta a_\mu$ are shown
in Figure~\ref{fig:chargeDiff}.
The
uncertainty of $\Delta a_\mu$ increases when the size of the charge is
doubled and, as expected, grows rapidly with decreasing quark
mass. Figure~\ref{fig:chargeDiff} shows
that the $\Delta Q=1/3 e$ and $\Delta Q=2/3 e$ QED corrections are highly
correlated, not unexpected because the same 
stochastic-wall source is used for 
the neutral and charged correlators.

The strange quark contribution makes up around $7\%$ of \amu.  An
advantage of computing the QED corrections to the connected strange
\amu, is that the larger mass of the strange quark compared to the
light quarks causes reduced errors for the strange \amu, so it is
potentially easier to determine the QED contribution. No chiral
extrapolation is required at the mass of the strange quark.

\begin{figure}[H]
    \centering
    \includegraphics[width=0.65\linewidth]{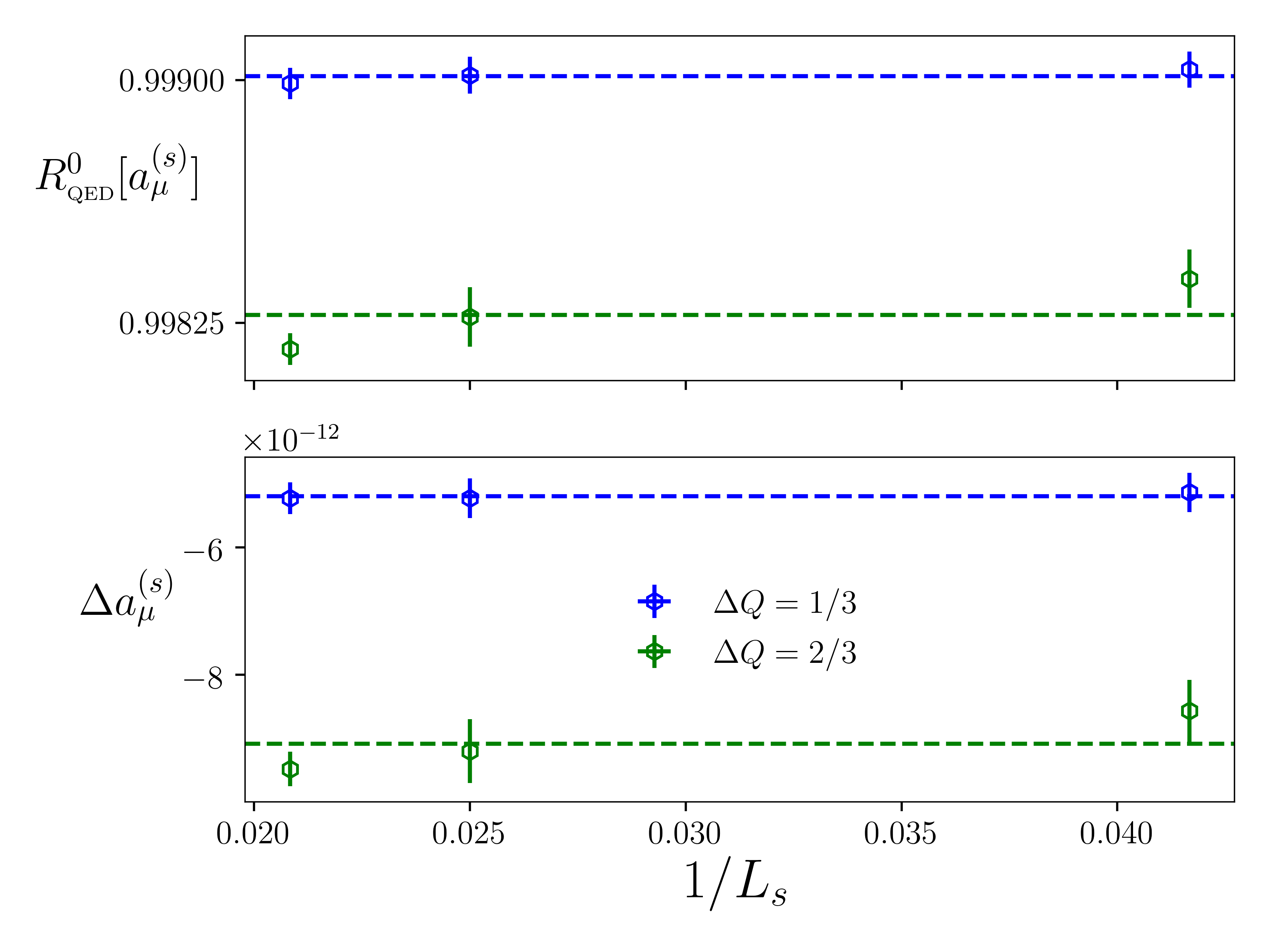}
    \caption{How $R_{\text{QED}}^0[a_\mu^{(s)}]$ and $\Delta
      a^{(s)}_{\mu}$ vary with inverse box size for $Q=-1/3e,2/3e$. The dashed lines are the mean of the three values.}
    \label{fig:FVE_amus}
\end{figure}

In order to assess the magnitude of possible QED finite volume effects
on $\Delta a^{(s)}_{\mu}$ we computed vector correlators on ensembles
with varying spatial volumes. Hatton et al.~\cite{Hatton:2020qhk} have
preformed a similar finite volume study for the charmonium $\Delta
a^{(c)}_{\mu}$.  In this study we used the coarse physical ensemble
with two ensembles with a
similar lattice spacing (0.12 fm) and unphysically heavy pions with
lattice volumes: $24^3 \times 64$ and $40^3 \times 64$.
We can define the ratio of a quantity with and without QED,
\begin{equation}
\begin{aligned}
  R^0_{\mbox{\tiny{QED}}}[X]\equiv \frac{X[\mbox{QCD+qQED}]}{X[\mbox{QCD}]}\hspace{5mm} \mbox{at fixed }am_s,
\end{aligned}
\end{equation}
and look at how this ratio varies with lattice
volume. Figure~\ref{fig:FVE_amus} shows how $\Delta a^{(s)}_\mu$ and
$R_{\text{QED}}^0[a_\mu^{(s)}]$ vary with the inverse lattice side
($1/L_s$). Figure~\ref{fig:FVE_amus} shows that the QED finite volume
effects on $\Delta a^{(s)}_{\mu}$ are negligible for the statistics
used. Even with the unphysical (for the strange quark) larger electric
charge $Q_s=-2/3 e$ the maximum deviation is only slightly above $1\sigma$.
The results of the finite volume study are similar to what was found for
charm quarks~\cite{Hatton:2020qhk} and is consistent with expectations from effective field theory~\cite{Bijnens:2019ejw}.


The QED contribution, $\delta a^{(s)}_{\mu}$, to $a^{(s)}_{\mu}$ is
obtained by taking the $Q=0$ and $Q=-1/3 e$ strange vector correlators
and computing $a^{(s)}_{\mu}$ and $\Delta a_\mu$, before 
converting to the Dashen scheme 
to obtain $\delta a_\mu$. We note that the scheme
adjustment on the fine ensemble is very imprecise,  because the
chiral extrapolation of the mass of the pseudoscalar meson to $m_l$
needed to get the quark mass shift is less constrained on
the fine ensemble. After the scheme adjustment the continuum limit
needs to be taken. As the HISQ action has lattice artifacts of
$\mathcal{O}(a^2)$ the data is fit to a simple function, linear in
$a^2$. The extrapolation
function used is:
\begin{equation}
    \delta a^{(s)}_{\mu}(a^2) = c_0 \left( 1 + c_1(a\Lambda)^2 \right),
    \label{eqn:strange_contX}
\end{equation}
where the $c_i$ are parameters to be fitted, $a$ is the lattice spacing, and 
$\Lambda$ =  0.5 GeV for the typical QCD scale.

\begin{figure}[H]
    \centering
    \includegraphics[width=0.65\linewidth]{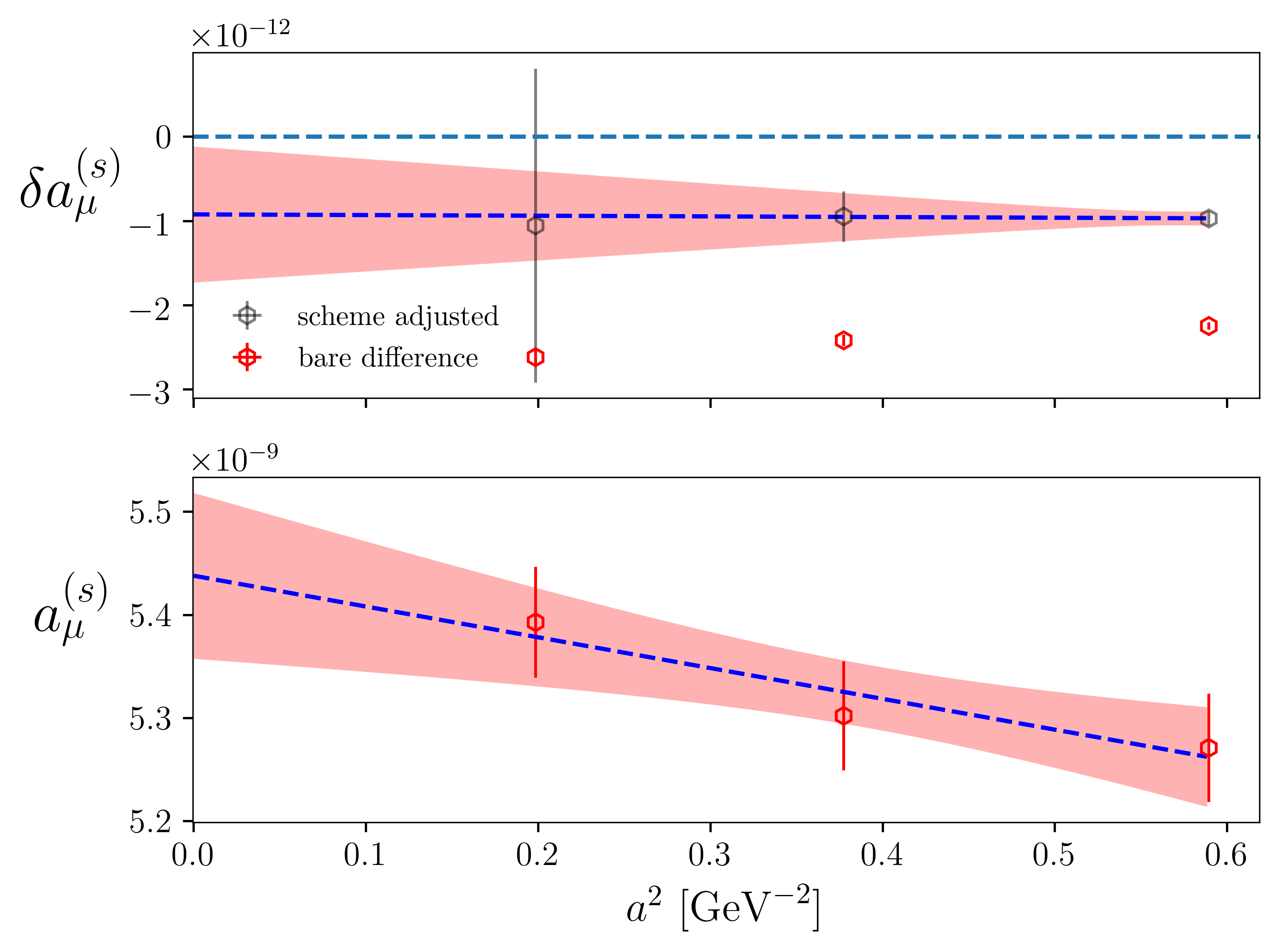}
    \caption{The blinded continuum extrapolations of $\delta a^{(s)}_{\mu}$ and $a^{(s)}_{\mu}$. See equation~\ref{eqn:strange_contX} and Table~\ref{tab:strangeX_fit} for the fit function and fitted parameter values.}
    \label{fig:strangeX}
\end{figure}

The extrapolation in $a^2$ is plotted in Figure~\ref{fig:strangeX} and
the fitted parameters are listed in Table~\ref{tab:strangeX_fit}. The
fit is satisfactory with goodness of fit parameters $\chi^2/{\rm
  dof}=0.21$ for the $a^{(s)}_{\mu}$ extrapolation and $\chi^2/{\rm
  dof}=0.0014$ for the $\delta a^{(s)}_{\mu}$
extrapolation. Figure~\ref{fig:strangeX} shows that the slope in $a^2$
is mild and the posterior for $c_1$ is consistent with a horizontal
band.

\begin{table}[H]
\centering
\begin{tabular}{ c c c }
\hline
param & prior & posterior \\
\hline
\multicolumn{3}{c}{$\delta a^{(s)}_{\mu}$} \\
\hline
$c_0$ & $0(1)\!\times\! 10^{-10}$ & $-0.0092(81)\!\times\! 10^{-10}$  \\
$c_1$ & 0(100) & $0.3(6.4)$ \\
\hline
\multicolumn{3}{c}{$a^{(s)}_{\mu}$} \\
\hline
$c_0$ & $0(1)\!\times\! 10^{-8}$ & $54.38(80)\!\times\! 10^{-10}$  \\
$c_1$ & $0(1)$ & $-0.22(14)$ \\
\hline
\end{tabular}
\caption{Priors and preliminary results for the parameters of the continuum extrapolations of $\delta a^{(s)}_{\mu}$ and $a^{(s)}_{\mu}$.}
\label{tab:strangeX_fit}
\end{table}

The extrapolated continuum values are,
\begin{equation}
\begin{aligned}
    \delta a^{(s)}_{\mu} &= -0.0092(81)\!\times\! 10^{-10} \\
    a^{(s)}_{\mu} &= 54.38(80)\!\times\! 10^{-10} \hspace{1mm},
\end{aligned}
\label{eqn:strange_result}
\end{equation}
where we remind the reader that these are blinded results. Our computed absolute uncertainty, $0.0081\times 10^{-10}$, contributes a tiny amount to the overall uncertainty of $a_{\mu}$.

The light quark contribution to $a_{\mu}$ comes from quark loops
formed from up and down quarks
$a^{(l)}_{\mu}$ makes up the lion's share, around
$90\%$, of the total value of $a_{\mu}$.
We work in the 
limit where the up and down quarks have the same mass (in
QCD), $m_l=\frac{1}{2}(m_u^{\text{phys}} +
m_d^{\text{phys}})$.  We measure vector correlators at
3$m_l$, 5$m_l$, 7$m_l$ for all ensembles, 
because measurements at the
physical pion mass are noisy.

\begin{figure}[H]
    \centering
    \includegraphics[width=0.9\linewidth]{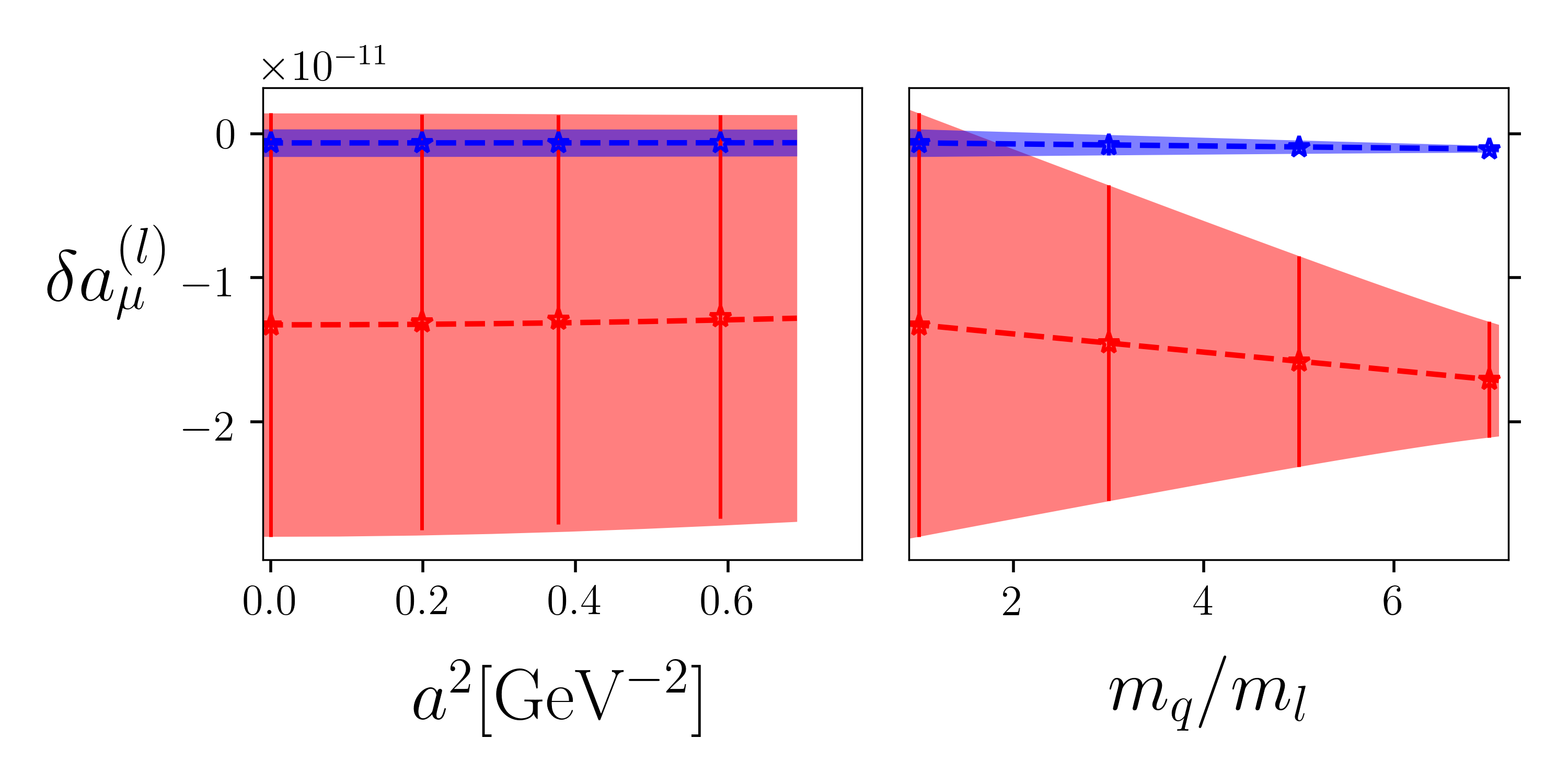}
    \caption{The blinded chiral-continuum extrapolations of $\delta
      a^{(u)}_{\mu}$ (red) and $\delta a^{(d)}_{\mu}$ (blue). On the left is the
      extrapolation in the lattice spacing at $m_q/m_l=1$ and on the
      right is the extrapolation in quark mass at $a=0$. See
      equations~\ref{eqn:light_combX} and
      Table~\ref{tab:lightX_fit}. }
    \label{fig:diff_continuum_ext}
\end{figure}

The procedure to obtain $\delta a^{(l)}_{\mu}$ is essentially the same
as that described above for $\delta a^{(s)}_{\mu}$. To aid the
calculation we split up $ \delta a^{(l)}_{\mu} = \delta a^{(u)}_{\mu}
+ \delta a^{(d)}_{\mu}$, where $\delta a^{(d)}_{\mu}$ is calculated using
the $Q=0,1/3e$ correlators and $\delta a^{(u)}_{\mu}$ is calculated
from the $Q=0,2/3 e$ correlators. To find the physical value of $\delta
a^{(l)}_{\mu}$ we do a combined chiral-continuum extrapolation for
each piece before adding them together. We fit our data to the
following functional form,
\begin{equation}
\begin{aligned}
    \delta a^{(d)}_{\mu}(a^2,m_q/m_l) &= c_0^{(d)}\left(1 + c_1^{(d)}(a\Lambda)^2 + c_2^{(d)} m_q/m_l \right)\\
    \delta a^{(u)}_{\mu}(a^2,m_q/m_l) &= c_0^{(u) }\left(1 + c_1^{(u)}(a\Lambda)^2 + c_2^{(u)} m_q/m_l \right),
    \end{aligned}
\label{eqn:light_combX}
\end{equation}
which has a similar form to the extrapolation used for the strange
quark contribution. The $c_2^{(u)} m_q/m_l$ and $c_2^{(d)} m_q/m_l$
terms control the extrapolation in quark mass. We again use
$\Lambda=0.5$ GeV. The fit has six parameters for 3 masses $\times$
3 ensembles $\times$ 2 (u/d) $=18$ pieces of data. We fit all six
parameters simultaneously to account for correlations between
measurements on the same ensemble.
We plot both the continuum and chiral extrapolations of $\delta
a^{(u)}_{\mu}, \delta a^{(d)}_{\mu}$ in
Figure~\ref{fig:diff_continuum_ext} and list the fitted values of the
$c_i$ in Table~\ref{tab:lightX_fit}.

\begin{table}[H]
\centering
\begin{tabular}{c c c }
\hline
param & prior & posterior \\
\hline
\multicolumn{3}{c}{$\delta a^{(u)}_{\mu}$} \\
\hline
$c_0$ & $0(1)\!\times\!10^{-9}$  & $-1.2(1.6)\!\times\!10^{-11}$ \\
$c_1$ & 0(1)      &  $-0.17(86)$ \\
$c_2$ & 0(1)      &  $0.06(23)$ \\
\hline
\multicolumn{3}{c}{$\delta a^{(d)}_{\mu}$} \\
\hline
$c_0$ & $0(1)\!\times\!10^{-10}$  & $-0.5(1.0)\!\times\!10^{-12}$ \\
$c_1$ & 0(1)      &  $-0.06(94)$ \\
$c_2$ & 0(1)      &  $0.14(51)$ \\
\hline
\end{tabular}
\caption{Priors and preliminary fitted results for the parameters of the chiral-continuum extrapolation of $\delta a^{(l)}_{\mu}$. See equations~\ref{eqn:light_combX} for fit function.}
\label{tab:lightX_fit}
\end{table}

Our extrapolated value for the quenched QED correction to the light connected HVP is
\begin{equation}
    \delta a^{(l)}_{\mu}(a^2=0,m_q/m_l=1) = -1.3(1.5)\times 10^{-11}  \hspace{1mm}.
\end{equation}
If the correlations between quark masses and charges are turned off we find an extrapolated $\delta a^{(l)}_{\mu}=1.9(3.4)\times 10^{-11}$.

\section{Conclusions}

We have used staggered quarks, gluon fields generated with the HISQ
action, and quenched U(1) fields gauge fixed with the QED$_\text{L}$
prescription to measure vector correlators at a series of lattice
spacings and light quark masses. From these correlators we have
computed the QED corrections to the light and strange 
connected \amu.

To compare our results on the QED contributions to \amu\ in the Dashen scheme
with those from other collaborations requires
work on converting the results to a consistent scheme.  Our results,
even though they are still blinded, imply that the
QED correction to the connected \amu\ are small, with an absolute
uncertainty less than $1\times 10^{-10}$. 
This is well below the
threshold necessary for sub percent precision on
the total $a_\mu$ ($\leq 5\times 10^{-10}$).
We are working on reducing the
errors on the other contributions to \amu\
to below sub percent precision,
which when combined
with the planned future experimental measurements of $a_\mu$ will
maximize the test of the theoretical prediction of $a_\mu$ from
the standard model~\cite{Aoyama:2020ynm}.

We also plan to study the contribution of QED to the windows on the
hadronic vacuum polarization~\cite{FermilabLattice:2022izv}, compute
the QED contributions to the disconnected diagrams and the effect of
QED in the sea.

\section*{Acknowledgements}

\noindent This work was supported in part by grants and contracts from the U.S. Department of Energy, Office of High Energy Physics, from the U.S. National Science Foundation, and from the Simons Foundation.

\bibliographystyle{JHEP}
\bibliography{add}

\providecommand{\href}[2]{#2}\begingroup\raggedright\begin{thebibliography}{10}

\bibitem{Muong-2:2021ojo}
{\scshape Muon {{\Large{g-}}2}} collaboration, \emph{{Measurement of the
  Positive Muon Anomalous Magnetic Moment to 0.46 ppm}},
  \href{https://doi.org/10.1103/PhysRevLett.126.141801}{\emph{Phys. Rev. Lett.}
  {\bfseries 126} (2021) 141801}
  [\href{https://arxiv.org/abs/2104.03281}{{\ttfamily 2104.03281}}].

\bibitem{Muong-2:2006rrc}
{\scshape {Muon {{\Large{g-}}2} }} collaboration, \emph{{Final Report of the
  Muon E821 Anomalous Magnetic Moment Measurement at BNL}},
  \href{https://doi.org/10.1103/PhysRevD.73.072003}{\emph{Phys. Rev. D}
  {\bfseries 73} (2006) 072003}
  [\href{https://arxiv.org/abs/hep-ex/0602035}{{\ttfamily hep-ex/0602035}}].

\bibitem{Aoyama:2020ynm}
T.~Aoyama et~al., \emph{{The anomalous magnetic moment of the muon in the
  Standard Model}},
  \href{https://doi.org/10.1016/j.physrep.2020.07.006}{\emph{Phys. Rept.}
  {\bfseries 887} (2020) 1} [\href{https://arxiv.org/abs/2006.04822}{{\ttfamily
  2006.04822}}].

\bibitem{Chakraborty:2014mwa}
{\scshape HPQCD} collaboration, \emph{{Strange and charm quark contributions to
  the anomalous magnetic moment of the muon}},
  \href{https://doi.org/10.1103/PhysRevD.89.114501}{\emph{Phys. Rev. D}
  {\bfseries 89} (2014) 114501}
  [\href{https://arxiv.org/abs/1403.1778}{{\ttfamily 1403.1778}}].

\bibitem{Chakraborty:2016mwy}
B.~Chakraborty, C.T.H.~Davies, P.G.~de~Oliviera, J.~Koponen, G.P.~Lepage and
  R.S.~Van~de Water, \emph{{The hadronic vacuum polarization contribution to
  $a_{\mu}$ from full lattice QCD}},
  \href{https://doi.org/10.1103/PhysRevD.96.034516}{\emph{Phys. Rev. D}
  {\bfseries 96} (2017) 034516}
  [\href{https://arxiv.org/abs/1601.03071}{{\ttfamily 1601.03071}}].

\bibitem{FermilabLattice:2017wgj}
{\scshape Fermilab Lattice, HPQCD, MILC} collaboration,
  \emph{{Strong-Isospin-Breaking Correction to the Muon Anomalous Magnetic
  Moment from Lattice QCD at the Physical Point}},
  \href{https://doi.org/10.1103/PhysRevLett.120.152001}{\emph{Phys. Rev. Lett.}
  {\bfseries 120} (2018) 152001}
  [\href{https://arxiv.org/abs/1710.11212}{{\ttfamily 1710.11212}}].

\bibitem{FermilabLattice:2019ugu}
{\scshape Fermilab Lattice, HPQCD, MILC} collaboration,
  \emph{{Hadronic-vacuum-polarization contribution to the
  muon\textquoteright{}s anomalous magnetic moment from four-flavor lattice
  QCD}}, \href{https://doi.org/10.1103/PhysRevD.101.034512}{\emph{Phys. Rev. D}
  {\bfseries 101} (2020) 034512}
  [\href{https://arxiv.org/abs/1902.04223}{{\ttfamily 1902.04223}}].

\bibitem{Lahert:2021xxu}
{\scshape Fermilab Lattice, HPQCD, MILC} collaboration, \emph{{Hadronic vacuum
  polarization of the muon on 2+1+1-flavor HISQ ensembles: an update.}},
  \href{https://doi.org/10.22323/1.396.0526}{\emph{PoS} {\bfseries LATTICE2021}
  (2022) 526} [\href{https://arxiv.org/abs/2112.11647}{{\ttfamily
  2112.11647}}].

\bibitem{FermilabLattice:2022izv}
{\scshape Fermilab Lattice, MILC, HPQCD} collaboration, \emph{{Windows on the
  hadronic vacuum polarization contribution to the muon anomalous magnetic
  moment}}, \href{https://doi.org/10.1103/PhysRevD.106.074509}{\emph{Phys. Rev.
  D} {\bfseries 106} (2022) 074509}
  [\href{https://arxiv.org/abs/2207.04765}{{\ttfamily 2207.04765}}].

\bibitem{MILC:2018ddw}
{\scshape MILC} collaboration, \emph{{Lattice computation of the
  electromagnetic contributions to kaon and pion masses}},
  \href{https://doi.org/10.1103/PhysRevD.99.034503}{\emph{Phys. Rev. D}
  {\bfseries 99} (2019) 034503}
  [\href{https://arxiv.org/abs/1807.05556}{{\ttfamily 1807.05556}}].

\bibitem{Fodor:2016bgu}
Z.~Fodor, C.~Hoelbling, S.~Krieg, L.~Lellouch, T.~Lippert, A.~Portelli et~al.,
  \emph{{Up and down quark masses and corrections to Dashen's theorem from
  lattice QCD and quenched QED}},
  \href{https://doi.org/10.1103/PhysRevLett.117.082001}{\emph{Phys. Rev. Lett.}
  {\bfseries 117} (2016) 082001}
  [\href{https://arxiv.org/abs/1604.07112}{{\ttfamily 1604.07112}}].

\bibitem{Follana:2006rc}
{\scshape HPQCD, UKQCD} collaboration, \emph{{Highly improved staggered quarks
  on the lattice, with applications to charm physics}},
  \href{https://doi.org/10.1103/PhysRevD.75.054502}{\emph{Phys. Rev. D}
  {\bfseries 75} (2007) 054502}
  [\href{https://arxiv.org/abs/hep-lat/0610092}{{\ttfamily hep-lat/0610092}}].

\bibitem{MILC:2010pul}
{\scshape MILC} collaboration, \emph{{Scaling studies of QCD with the dynamical
  HISQ action}}, \href{https://doi.org/10.1103/PhysRevD.82.074501}{\emph{Phys.
  Rev. D} {\bfseries 82} (2010) 074501}
  [\href{https://arxiv.org/abs/1004.0342}{{\ttfamily 1004.0342}}].

\bibitem{MILC:2012znn}
{\scshape MILC} collaboration, \emph{{Lattice QCD Ensembles with Four Flavors
  of Highly Improved Staggered Quarks}},
  \href{https://doi.org/10.1103/PhysRevD.87.054505}{\emph{Phys. Rev. D}
  {\bfseries 87} (2013) 054505}
  [\href{https://arxiv.org/abs/1212.4768}{{\ttfamily 1212.4768}}].

\bibitem{Dowdall:2013rya}
R.J.~Dowdall, C.T.H.~Davies, G.P.~Lepage and C.~McNeile, \emph{{Vus from pi and
  K decay constants in full lattice QCD with physical u, d, s and c quarks}},
  \href{https://doi.org/10.1103/PhysRevD.88.074504}{\emph{Phys. Rev. D}
  {\bfseries 88} (2013) 074504}
  [\href{https://arxiv.org/abs/1303.1670}{{\ttfamily 1303.1670}}].

\bibitem{Duncan:1996xy}
A.~Duncan, E.~Eichten and H.~Thacker, \emph{{Electromagnetic splittings and
  light quark masses in lattice QCD}},
  \href{https://doi.org/10.1103/PhysRevLett.76.3894}{\emph{Phys. Rev. Lett.}
  {\bfseries 76} (1996) 3894}
  [\href{https://arxiv.org/abs/hep-lat/9602005}{{\ttfamily hep-lat/9602005}}].

\bibitem{Hatton:2020qhk}
{\scshape HPQCD} collaboration, \emph{{Charmonium properties from lattice
  $QCD$+QED : Hyperfine splitting, $J/\psi$ leptonic width, charm quark mass,
  and $a^c_\mu$}},
  \href{https://doi.org/10.1103/PhysRevD.102.054511}{\emph{Phys. Rev. D}
  {\bfseries 102} (2020) 054511}
  [\href{https://arxiv.org/abs/2005.01845}{{\ttfamily 2005.01845}}].

\bibitem{Hayakawa:2008an}
M.~Hayakawa and S.~Uno, \emph{{QED in finite volume and finite size scaling
  effect on electromagnetic properties of hadrons}},
  \href{https://doi.org/10.1143/PTP.120.413}{\emph{Prog. Theor. Phys.}
  {\bfseries 120} (2008) 413}
  [\href{https://arxiv.org/abs/0804.2044}{{\ttfamily 0804.2044}}].

\bibitem{Hatton:2019gha}
{\scshape HPQCD} collaboration, \emph{{Renormalizing vector currents in lattice
  QCD using momentum-subtraction schemes}},
  \href{https://doi.org/10.1103/PhysRevD.100.114513}{\emph{Phys. Rev. D}
  {\bfseries 100} (2019) 114513}
  [\href{https://arxiv.org/abs/1909.00756}{{\ttfamily 1909.00756}}].

\bibitem{Borsanyi:2020mff}
S.~Borsanyi et~al., \emph{{Leading hadronic contribution to the muon magnetic
  moment from lat tice QCD}},
  \href{https://doi.org/10.1038/s41586-021-03418-1}{\emph{Nature} {\bfseries
  593} (2021) 51} [\href{https://arxiv.org/abs/2002.12347}{{\ttfamily
  2002.12347}}].

\bibitem{Bali:2009hu}
G.S.~Bali, S.~Collins and A.~Sch{\"a}fer, \emph{{Effective noise reduction
  techniques for disconnected loops in Lattice QCD}},
  \href{https://doi.org/10.1016/j.cpc.2010.05.008}{\emph{Comput. Phys. Commun.}
  {\bfseries 181} (2010) 1570}
  [\href{https://arxiv.org/abs/0910.3970}{{\ttfamily 0910.3970}}].

\bibitem{Bijnens:2019ejw}
J.~Bijnens, J.~Harrison, N.~Hermansson-Truedsson, T.~Janowski, A.~J\"uttner and
  A.~Portelli, \emph{{Electromagnetic finite-size effects to the hadronic
  vacuum polarization}},
  \href{https://doi.org/10.1103/PhysRevD.100.014508}{\emph{Phys. Rev. D}
  {\bfseries 100} (2019) 014508}
  [\href{https://arxiv.org/abs/1903.10591}{{\ttfamily 1903.10591}}].

\end{thebibliography}\endgroup

\end{document}